# Second-harmonic assisted four-wave mixing in chip-based microresonator frequency comb generation

Xiaoxiao Xue[1,2*], François Leo[3,4], Yi Xuan[2,5], Jose A. Jaramillo-Villegas[2,6], Pei-Hsun Wang[2], Daniel E. Leaird[2], Miro Erkintalo[3], Minghao Qi[2,5], and Andrew M. Weiner[2,5†]

[1]Department of Electronic Engineering, Tsinghua University, Beijing 100084, China
[2]School of Electrical and Computer Engineering, Purdue University, 465 Northwestern Avenue, West Lafayette, Indiana 47907-2035, USA
[3]The Dodd-Walls Centre for Photonic and Quantum Technologies, Department of Physics, The University of Auckland, Auckland 1142, New Zealand
[4]OPERA-photonics, Université libre de Bruxelles (U.L.B.), 50 Avenue F. D. Roosevelt, CP 194/5, B-1050 Bruxelles, Belgium
[5]Birck Nanotechnology Center, Purdue University, 1205 West State Street, West Lafayette, Indiana 47907, USA
[6]Facultad de Ingenierías, Universidad Tecnológica de Pereira, Pereira, RIS 660003, Colombia
*xuexx@tsinghua.edu.cn; †amw@purdue.edu

**ABSTRACT**

Simultaneous Kerr comb formation and second-harmonic generation with on-chip microresonators can greatly facilitate comb self-referencing for optical clocks and frequency metrology. Moreover, the presence of both second- and third-order nonlinearities results in complex cavity dynamics that is of high scientific interest but is still far from well understood. Here, we demonstrate that the interaction between the fundamental and the second-harmonic waves can provide an entirely new way of phase-matching for four-wave mixing in optical microresonators, enabling the generation of optical frequency combs in the normal dispersion regime, under conditions where comb creation is ordinarily prohibited. We derive new coupled time-domain mean-field equations and obtain simulation results showing good qualitative agreement with our experimental observations. Our findings provide a novel way of overcoming the dispersion limit for simultaneous Kerr comb formation and second-harmonic generation, which might prove especially important in the near-visible to visible range where several atomic transitions commonly used for stabilization of optical clocks are located and where the large normal material dispersion is likely to dominate.

## INTRODUCTION

Comb-based optical frequency metrology and atom clocks have pushed the measurement of time and frequency to an unprecedented precision in past decades [1]. Recently emerged miniature microresonator-based optical frequency combs (microcombs) can potentially bring this technology from lab to portable applications [2]−[10]. Microcomb generation is a technique that converts a single-frequency laser pump to a broadband comb source using a high-quality-factor microresonator. The basic mechanism is based on the cavity enhanced Kerr effect arising from the third-order optical nonlinearity. Rich dynamics in the comb generation process has been observed, such as chaos [11], mode-locking [12]−[17], temporal bright and dark solitons [18]−[22], discrete phase steps [23], Cherenkov radiation [24], etc. Besides ultra-wideband comb generation, second-harmonic conversion of a fraction of the frequency lines is another key step to achieve a fully stabilized comb source based on comb self-referencing. Simultaneous comb formation and second-harmonic generation with a single microresonator will greatly facilitate this procedure because it avoids the need of additional second-order nonlinear elements. This can potentially be achieved by taking advantage of both second- and third-order nonlinearities in some microresonator platforms [25],[26]. Nevertheless, compared to a purely third-order system, the participation of second-order nonlinearity can affect the comb self-starting and mode-locking behaviors, resulting in complex dynamics that is far from well understood.

In this work, we report a nonlinear mode coupling mechanism which can significantly change the comb dynamics in microresonators with both second- and third-order nonlinearities. The mode coupling arises from the interaction between the fundamental and the second-harmonic waves, and permits modulational instability (MI) in the normal-dispersion region under conditions where comb formation is typically prohibited in the absence of mode coupling. In both experiments and numerical simulations, we demonstrate simultaneous second-harmonic generation and broadband mode-locked comb generation in a microresonator which shows normal group velocity dispersion for the fundamental wave. Our work reveals, for the first time to our knowledge, that second-harmonic mode coupling can provide an entirely new way of fulfilling phase-matching for four-wave mixing in the normal dispersion regime, and may also alter the nonlinear dynamics in the anomalous dispersion regime.

## MATERIALS AND METHODS

### Silicon nitride microring fabrication

The microresonator we use is a silicon nitride (SiN) microring embedded in silicon dioxide; this is one of the microrings used in our previous report on mode-locked dark pulses [20]. The microring has a radius of 100 μm corresponding to a free spectral range (FSR) of ~231 GHz, and a loaded quality factor of $8.6 \times 10^5$. The cross-section of the microring waveguide is 2 μm $\times$ 550 nm, and its measured group velocity dispersion in the C-band is 186.9 $ps^2$/km, i.e., strongly normal. Both a through-port waveguide and a drop-port waveguide coupled to the microring are fabricated. The device fabrication process is as follows. An under cladding layer of 3 μm thermal oxide is grown on a silicon wafer in an oxidation tube at 1100 °C. Using low-pressure chemical vapor deposition, a 550 nm SiN film is deposited at 800 °C on the oxidized wafer. A negative hydrogen silsesquioxane (HSQ) resist is used to pattern the waveguide and resonator via an electron beam lithography system at 100 kV. After developing in tetramethylammonium hydroxide solution, the HSQ pattern is transferred to the SiN film using reactive ion etching. Then, a 3.5 μm thick low-temperature oxide film, which serves as an upper cladding, is deposited at 400 °C followed by an annealing step undertaken at 1100 °C in an $N_2$ atmosphere.

### Comb generation and characterization

A detailed schematic of our experimental setup for comb generation is shown in Fig. 1(a). An infrared (IR) tunable laser source in the C-band is amplified and pumps the SiN microring. The light from the microring chip is split into three branches. From top to bottom, the first branch measures the total power in the fundamental IR spectral range; the second branch measures the IR comb power excluding the pump line (the pulse shaper acts as a band-reject filter); the third branch measures the second-harmonic power (the IR power is blocked by a glass filter). The Pound–Drever–Hall (PDH) signal is also measured to monitor the effective pump-resonance detuning when the optical signal is analyzed at the through port [18] (setup not shown).

To probe the phase coherence of the generated frequency comb and to investigate its time-domain characteristics, we have used spectral line-by-line shaping in combination with intensity correlation measurements [12],[13]. The dispersion of the fiber link between the output of the microring chip and the input of the intensity correlator is compensated; thus we can retrieve the comb spectral phase by compressing the comb to a transform-limited pulse. The time-domain waveform can then be reconstructed through Fourier synthesis based on the amplitude and phase information of the comb. Self-referenced cross-correlation is also measured for comparison.

In diagnosing the comb formation dynamics, we have introduced a method of fiber comb spectroscopy to measure the relative positions of the microresonator resonances in comb operation with respect to the comb lines (i.e. hot-cavity detuning). The principle will be explained in detail together with the results in a later section.

**Theoretical model**

In microresonators with only third-order Kerr nonlinearity, the evolution of the intracavity field can generally be described by the mean-field Lugiato–Lefever (L-L) equation [27],[28]. When there are simultaneous second- and third-order nonlinearities, a second-harmonic wave may be generated, with the fundamental and second-harmonic waves coupled to each other through second-order nonlinearity. To understand the underlying physics for our experiments, a theoretical model incorporating coupled L-L equations is developed (see Supplementary Section 2 for the equation derivation [29],[30]). In particular, we find that the intracavity fields at the fundamental and the second-harmonic wavelengths obey the following coupled equations:

$$\frac{\partial E_1}{\partial z} = \left[ -\alpha_1 - \mathrm{i}\delta_1 - \mathrm{i}\frac{k_1^{"}}{2}\frac{\partial^2}{\partial \tau^2} + \mathrm{i}\gamma_1 |E_1|^2 + \mathrm{i}2\gamma_{12} |E_2|^2 \right] E_1 + \mathrm{i}\kappa E_2 E_1^* + \eta_1 E_{\mathrm{in}} \qquad (1)$$

$$\frac{\partial E_2}{\partial z} = \left[ -\alpha_2 - \mathrm{i}2\delta_1 - \mathrm{i}\Delta k - \Delta k' \frac{\partial}{\partial \tau} - \mathrm{i}\frac{k_2^{"}}{2}\frac{\partial^2}{\partial \tau^2} + \mathrm{i}\gamma_2 |E_2|^2 + \mathrm{i}2\gamma_{21} |E_1|^2 \right] E_2 + \mathrm{i}\kappa^* E_1^2 \qquad (2)$$

Here, $E_1$ and $E_2$ are the intracavity amplitudes of the fundamental and second-harmonic waves scaled such that $|E_{1,2}|^2$ represents the power flow; $z$ is propagation distance in the cavity; $\tau$ is time; $\alpha_1$ and $\alpha_2$ are averaged loss per unit length including intrinsic loss and external coupling loss; $\delta_1$ is related to the pump-resonance detuning by $\delta_1 = (\omega_0 - \omega_\mathrm{p})t_\mathrm{R}/L$, with $\omega_0$ resonance frequency, $\omega_\mathrm{p}$ pump frequency, $t_\mathrm{R}$ round-trip time for the fundamental wave, $L$ round-trip length; $k_1^{"} = \mathrm{d}^2k/\mathrm{d}\omega^2\big|_{\omega=\omega_\mathrm{p}}$, $k_2^{"} = \mathrm{d}^2k/\mathrm{d}\omega^2\big|_{\omega=2\omega_\mathrm{p}}$ group velocity dispersion; $\Delta k = 2k(\omega_\mathrm{p}) - k(2\omega_\mathrm{p})$ phase mismatch; $\Delta k' = \mathrm{d}k/\mathrm{d}\omega\big|_{\omega=2\omega_\mathrm{p}} - \mathrm{d}k/\mathrm{d}\omega\big|_{\omega=\omega_\mathrm{p}}$ group velocity mismatch; $\gamma_1$, $\gamma_2$ nonlinear coefficients of self-phase modulation; $\gamma_{12}$, $\gamma_{21}$ nonlinear coefficients of cross-phase modulation; $\kappa$ second-order coupling coefficient; $\eta_1 = \sqrt{\theta_1}/L$ coupling coefficient between the pump and the intracavity field where $\theta_1$ is waveguide-resonator power coupling ratio for the fundamental wave; $E_{\mathrm{in}}$ amplitude of the pump.

## RESULTS AND DISCUSSION

### Second-harmonic assisted comb generation

Although second-order nonlinearity is absent in bulk amorphous SiN material, it can exist in SiN thin films and waveguides due to surface effects or due to silicon nanocrystals formed in the fabrication process [31]−[35]. Indeed, in our experiments we find that selected cavity resonances yield significant generation of light at the second-harmonic of the pump laser. To generate a frequency comb, the wavelength of the pump laser is tuned around 1543 nm. Specifically, the laser sweeps from shorter wavelength to longer wavelength across the microring resonance at a relatively slow speed of 0.5 nm/s. The on-chip pump power is around 0.4 W. Figure 1(b) shows the transmission curves measured at the output of the microring chip, with the zoom-in details shown in Fig. 1(c). Due to the resonance shift caused by the Kerr and thermo-optic effects, the total power transmission has a typical triangular shape [36]. Furthermore, a signal at the second-harmonic can readily be detected as the pump laser tunes into resonance. Transition steps, which correspond to switching of the comb dynamics, can also be clearly observed in transmission curves corresponding to both the fundamental and the second-harmonic wavelengths. The PDH signal shown in the bottom panel of Fig. 1(b) indicates that the pump laser in comb operation is always effectively blue detuned with respect to the shifted resonance.

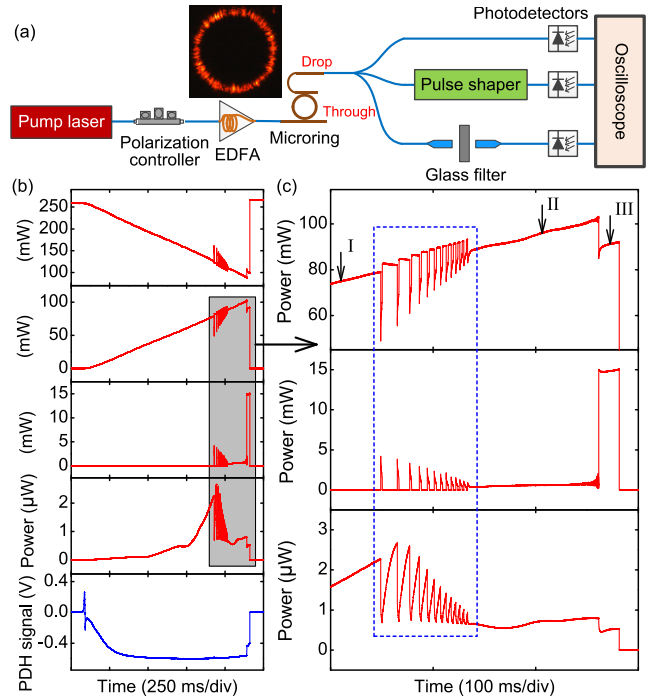

**Figure 1** Second-harmonic assisted comb generation in a normal-dispersion silicon nitride microring resonator. (a) Experimental setup. The inset shows the microscope image of the microring with second-harmonic radiation. (b) Transition curves when the pump laser scans across the resonance from shorter to longer wavelength. From top to bottom: total infrared (IR) power at the through port, total IR power at the drop port, IR comb power excluding the pump line at the drop port, second-harmonic power at the drop port, Pound–Drever–Hall (PDH) signal (the measurement setup is not shown in (a)). (c) Zoom-in details in the gray region marked in (b). The region marked in a blue dashed box shows a millisecond time scale oscillation. The drop-port comb spectra at different detuning stages (I, II, III) are shown in the following Fig. 2.

The optical power shows an oscillation in the region marked with a blue dashed box in Fig. 1(c). The oscillation occurs on a millisecond time scale and increases in speed as the pump laser is tuned to the red. Oscillation and self-pulsing caused by thermal nonlinearity has been observed in a variety of microresonator platforms [37]−[42]. It generally requires either two nonlinear mechanisms with different signs or two excited modes that interact with each other [37],[38]. For our microring, although two nonlinear processes with different response times were observed previously in microheater actuated tuning experiments [43], the two processes appear to have the same sign. We believe the oscillation may involve the interplay between the pump and comb modes with a resonant mode at the second-harmonic frequency. As supporting evidence, we note that (i) switching of the comb power and second-harmonic power are both observed within the oscillation in Fig. 1(c), and (ii) we did not see the thermal oscillations when pumping resonances of the same microring that do not show second-harmonic generation – including those where we previously observed linear-mode-coupling initiation of the comb. In the current paper, we focus our investigations in the region where the oscillation vanishes with the pump laser further tuned to the red, and leave the development of a detailed oscillation model for future work.

The optical spectra measured in a range spanning more than one octave at different pump detunings are shown in Fig. 2(a) (the increase on the lowest frequency side is due to the optical spectrum analyzer background, not comb lines). The zoom-in spectra of the fundamental and second-harmonic waves are shown in Figs. 2(b) and 2(c) respectively. Frequency combs can clearly be observed both around the IR pump as well as its second-harmonic. The comb formation can be divided into three different stages as follows. At detuning stage I (see Fig. 1(c)), only the second harmonic of the pump can be observed and no comb is generated. With the detuning changed to stage II, a narrow-band fundamental IR comb spaced by 1 FSR is generated and a few second-harmonic lines show up. The spacing between the second-harmonic lines is equal to that of the fundamental comb. The IR comb then transitions to a state with a much broader spectrum at detuning stage III. A slight change of the second-harmonic spectra is also observed in the transition. The IR comb shows very low intensity noise (below the noise background of the measuring electrical spectrum analyzer) in most part of the pump sweeping process, except in a small region right before transitioning to stage III where an increased intensity noise is observed. (The intensity noise is not revealed in the transmission curves in Fig. 1(c) due to a smoothing function of the oscilloscope; see Supplementary Section 1 for traces captured with a different data acquisition mode.) Note that the low intensity noise of the broadband comb at stage III generally corresponds to mode-locking and formation of temporal structures in the cavity [15]−[20].

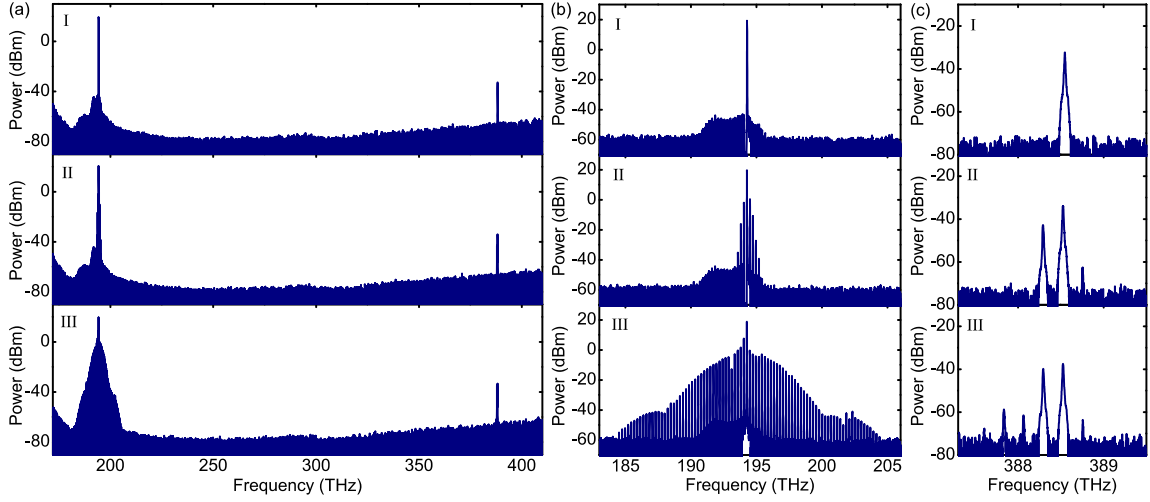

**Figure 2** Drop-port comb spectra at different detuning stages marked in Fig. 1(c). (a) Spectra measured in a range more than one octave (the increase on the lowest frequency side is due to the optical spectrum analyzer background). (b) Zoom-in infrared (IR) spectra. (c) Zoom-in second-harmonic spectra.

**Reciprocal bright and dark pulses**

Figures 3(a) and 3(b) show respectively the amplitude and phase of the comb lines in the C-band at stage III (here the comb phase follows the sign convention generally used in ultrafast optics [44], i.e. the field of each comb line is represented by $Ae^{i(\omega t+\phi)}$ where $A$ is the amplitude, $\omega$ is the angular frequency, and $\phi$ is the phase). The only difference between the optical fields at the through and drop ports is the complex amplitude of the pump line. All the comb lines at the drop port arise from coupling out of the microresonator. Therefore, the complex spectrum at the drop port should be the same as that in the cavity, as should the time-domain waveform. However, the pump line at the through port corresponds to the coherent superposition of the pump transmitted directly from the input waveguide and the pump component coupled out of the cavity. Because the complex pump field at the through port differs from that in the cavity, the time-domain waveforms at the through port and inside the cavity also differ [20]. For the specific example described here, the pump lines at the through and drop ports have very close power levels (difference <1 dB), but have a large phase difference (~2.4 rad). Figures 3(c) and 3(d) show the reconstructed waveforms at the through and drop ports respectively. The drop-port (intracavity) waveform is a dark pulse, which has been previously reported in normal-dispersion microresonators [20]. The dark pulse width is ~394 fs. Interestingly, the through-port waveform shows a nice bright pulse (width ~275 fs). Note that the bright pulse has a high background level which is limited by the cancellation ratio between the input pump and the dark pulse background in the through-port waveguide. The background level can be adjusted by changing the coupling condition between the microresonator and the through-port waveguide. Close-to-zero background can be achieved for the critical coupling condition. This way to tailor the pulse shape may be potentially useful for applications that require ultrashort bright pulses. Self-referenced cross-correlation is also measured to further verify the time-domain waveforms [20], and the results are consistent with those reconstructed through spectral line-by-line shaping (see Figs. 3(e) and 3(f)).

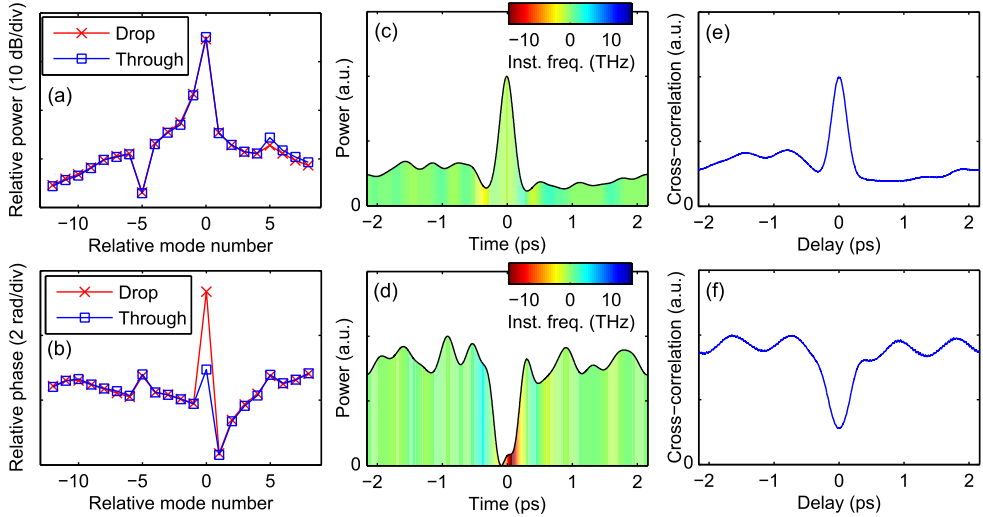

**Figure 3** Time-domain characterization of the broadband mode-locked comb at stage III. (a), (b) Amplitude and phase of the comb lines in the C-band. The amplitude of the pump line is very close at the through and drop ports, while a clear phase difference can be observed. (c), (d) Reconstructed waveforms at the through and drop ports, showing bright and dark pulses respectively. Inst. freq.: instantaneous frequency. (e), (f) Results of self-referenced cross-correlation which are consistent with (c), (d).

## Hot-cavity detuning measurement

The normal dispersion microring we investigate here is the same ring we used in our previous paper [20]. In that work we showed that (i) the mode spectrum of the microring is perturbed by avoided mode crossings induced by linear coupling between transverse modes, and that (ii) the phase shifts associated with the linear coupling can enable comb generation, even in the normal dispersion regime, when pumping resonances around a mode crossing region. This mechanism cannot, however, explain the comb generation described here in the current paper. Indeed, a clear signature of linear mode coupling induced comb generation is that one of the initial comb sideband pair always grows from the resonance which is shifted from its natural frequency due to mode coupling [45]−[47]. In contrast, here the laser pumps a resonance 4-FSR (~7 nm) away from the mode crossing region; no frequency shifts due to mode coupling are observed for the resonances close to the pump from the cold-cavity dispersion profile, nor are linear mode coupling induced comb lines observed anywhere in the greater than one octave spectral measurement range. Of course, the mode crossing position may slightly shift in a hot cavity when the microresonator is pumped because different transverse modes may have different thermal shifting rates [20]. Thus to provide strong evidence that the comb generation we observe in Fig. 2 is not due to linear mode coupling in the hot cavity, the relative positions of the resonances in comb operation with respect to the comb lines (i.e. hot-cavity detuning) are measured.

The hot-cavity detuning can be measured by using weakly modulated light [48],[49],[23]. In Ref. [23], a probe laser passed through the cavity is used to measure the resonance transmission in the backward direction; and the relative position of the corresponding comb line can be resolved from its beat note with the probe laser. But this method is not suitable for our setup due to a strong backward reflection of the comb power from the waveguide facet. Here we use an improved method based on fiber comb spectroscopy. The setup is shown in Fig. 4(a). A fiber frequency comb with a 20-MHz repetition rate is injected into the microring from the drop port via a circulator. The power of the fiber comb is adjusted such that it does not introduce any observable thermal and Kerr shifting. The fiber comb lines passing through the microring are gathered via another circulator at the pump input port. Meanwhile a fraction of the microcomb power is reflected to the backward direction from the waveguide facet and goes together with the fiber comb through the circulator (see the illustration in Fig. 4(a)). The fiber comb and microcomb lines from the circulator are combined with a tunable reference laser. Their radiofrequency (RF) beat notes are detected with a photodetector and analyzed with an electrical spectrum analyzer. To measure a specified resonance and the corresponding microcomb line, the frequency of the reference laser is tuned ~10 GHz lower than that of the microcomb line. The fiber comb lines pick up the shape of the microring transmission spectrum and then beat with the reference laser. The relative resonant frequency is obtained by fitting the intensities of the fiber comb lines with a Lorentzian function, while the relative frequency location of the microcomb line is determined directly from its beat note with the reference laser. The hot-cavity detuning, defined as the resonance frequency minus the pump (or comb line) frequency, can then be calculated as the resonance RF beat frequency minus the microcomb line RF beat frequency. Two example RF beat spectra indicating different detunings are shown in Fig. 4(b). The measurement accuracy of this method is estimated to be at the MHz level which is much smaller than the resonance width (~200 MHz).

The measured hot-cavity detuning results are shown in Fig. 4(c). All the comb lines including the pump are always blue detuned with respect to the resonances (i.e., resonance frequency < frequency of pump or comb line, respectively). The detuning value is even larger than the resonance width. At stage III, the detuning versus wavelength shows as a quadratic function which indicates a normal group velocity dispersion of ~190 $ps^2$/km, equal within experimental error to the dispersion measured for the cold cavity. The jump around 1535 nm is due to linear transverse mode coupling. Note that no detuning jumps are observed for the two resonances adjacent to the pump at either stage II or stage III. This is consistent with our conclusion that the 1-FSR comb at stage II in Fig. 2(b) is not due to linear mode coupling. At this point, we also note that the formation of the comb at stage II cannot be attributed to the usual MI induced by cavity boundary conditions in normally dispersive cavities [50]−[52],[21],[22]: that instability only exists in a region that is not accessed when the pump laser is tuned from blue to red into the resonance [27],[53],[54], as we do in our experiments. The comb formation dynamics is not a complete analogy to that of self-induced MI lasers either [55], since the microresonator we use is a passive cavity without gain medium.

Another interesting feature is that the hot-cavity detuning at stage II shows a linear slope versus wavelength superimposed onto the same quadratic function from stage III, suggesting a mismatch between the comb line spacing and the cavity FSR. To further verify this behavior, the comb line spacing is measured by using electro-optic down-mixing [56] and the results are shown in Fig. 4(d). The comb line spacing is changed by 212.7 MHz from stage II to stage III which agrees with the number retrieved from the detuning slope (214 MHz). Remarkably, all the key characteristics observed in experiments, i.e. the transition in the comb spectrum, as well as the abrupt change in comb line spacing, are accurately captured by numerical simulations based on a theoretical model that includes the second-order nonlinear interactions between the fundamental and second-harmonic waves. In particular, these simulations unequivocally reveal that the comb formation is triggered by the fundamental−second-harmonic interactions. The details of these simulations will be shown in the next section.

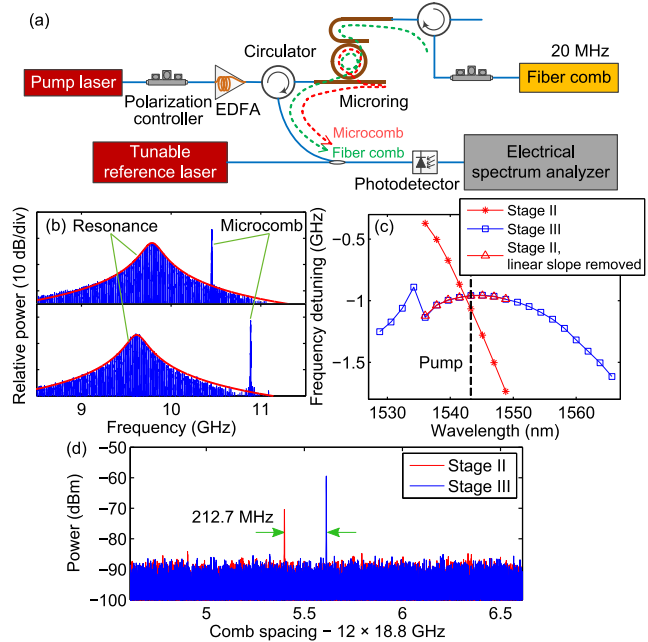

**Figure 4** Hot-cavity detuning measurement. (a) Experimental setup. The frequency of the reference laser is ~10 GHz lower than the microcomb line to be measured. (b) Example beat notes observed on the electrical spectrum analyzer. (c) Hot-cavity detuning defined as the resonance frequency minus the microcomb frequency. A change of the linear slope can be observed when the comb transitions from stage II to stage III, indicating a jump of the comb line spacing. (d) Measured microcomb beat note with electro-optic down-mixing, showing the comb line spacing jump.

## Numerical simulation

As mentioned above, MI is absent in a microresonator with purely Kerr nonlinearity when the group velocity dispersion is normal and when the pump laser is tuned from blue to red into the resonance as we do in our experiments [27]; thus no comb generation should be observed. (Although linear mode coupling may enable comb generation [45]−[47], as we have explained above, our observations in the current experiment cannot be explained by this mechanism.) However, when there are simultaneous second- and third-order nonlinearities, a second-harmonic wave may be generated, with the fundamental and second-harmonic waves coupled to each other through second-order nonlinearity. The evolution of the intracavity fields in this case follows coupled L-L equations (see Eqs. (1) and (2) above) rather than the single L-L equation generally considered in Kerr comb generation. New dynamics may then arise which are responsible for the comb generation described in the current paper.

    Figure 5 shows the simulation results based on Eqs. (1) and (2), which are very similar to our experimental observations (see Supplementary Section 2.2 for simulation parameters and Section 2.3 for a side-by-side plot of the experimental and simulation results). The thermo-optic effect is not considered in the simulation. With the pump laser continuously tuned into the

resonance from the blue side, a narrow-band fundamental comb is first generated, and then transitions to a much broader spectrum which corresponds to temporal dark pulse formation in the cavity. Detailed MI analyses (Supplementary Section 2.2) show that, due to the second-order nonlinear interactions, the frequency modes of the fundamental wave are coupled to those of the second-harmonic wave, resulting in an effect similar to linear transverse mode coupling. Specifically, in the simulation example shown in Fig. 5, a close-to-phase-matching condition is achieved such that one comb sideband mode 1-FSR away from the pump is strongly coupled to its sum frequency with the pump through sum/difference frequency generation. The mode interaction gives rise to phase shifts that result in equivalent anomalous dispersion for the pump and the two adjacent fundamental modes. MI can thus occur, giving rise to the generation of a 1-FSR comb at stage II. The comb state becomes unstable in a small region right before transitioning to stage III, which is consistent with the experimental observation (see Supplementary Section 1). At this point we emphasize that, as expected based on the strong normal dispersion, no modulation instability (or comb formation) is observed in our simulations when the second-order nonlinearity is neglected ($\kappa = 0$). This further confirms that it is precisely the nonlinear mode coupling that permits comb generation for our parameters.

The evolution of the fundamental intracavity waveform with the propagation distance is shown in Fig. 6. Note that the waveform at stage II shows a time drift which implies a deviation between the comb line spacing and the cavity FSR. The shifting rate is $\Delta \tau = 5.04$ fs per round trip. The abrupt shift of the comb line spacing from stage II to stage III is then estimated as $\Delta \tau \cdot FSR^2 \approx 270$ MHz which is very close to the experimental result (212.7 MHz). The mismatch between the comb line spacing and the cavity FSR is caused by the second-harmonic mode coupling which introduces additional phase shifts to the comb lines when the fundamental wave propagates in the cavity. The additional phase shifts effectively shift the resonances seen by the comb lines resulting in an equivalent change of the local FSR. This effect is prominent at stage II (which achieves an effective anomalous dispersion for MI), and gets weaker at stage III due to the transition to the broader mode-locked comb state.

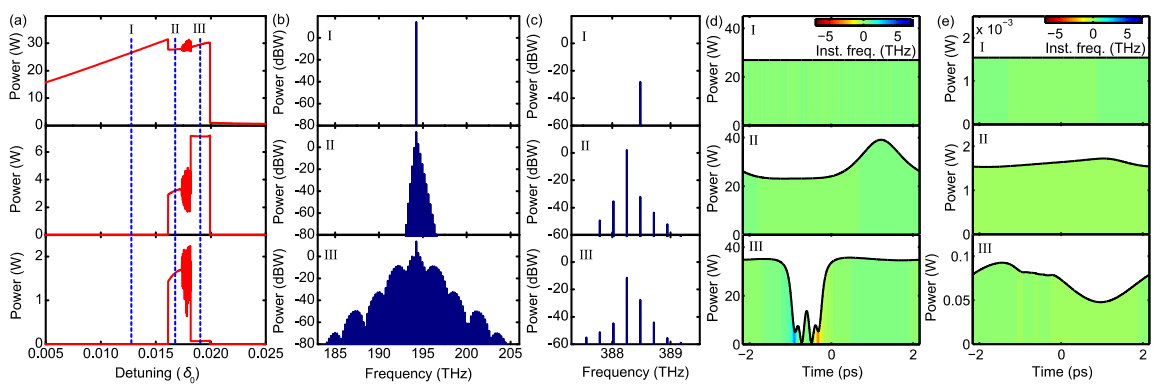

**Figure 5** Numerical simulation based on coupled L-L equations. (a) Intracavity power versus detuning ($\delta_0 = (\omega_0 - \omega_p)t_R$). From top to bottom: total fundamental power, fundamental comb power excluding the pump, total second-harmonic power. (b) Spectra of the fundamental wave at different detuning stages marked in (a). (c) Spectra of the second-harmonic wave. (d) Time-domain waveforms of the fundamental wave. Inst. freq.: instantaneous frequency. (e) Time-domain waveforms of the second-harmonic wave.

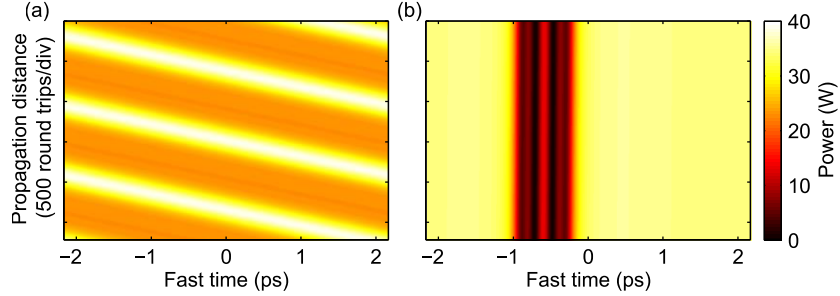

**Figure 6** Evolution of the intracavity fundamental wave with the propagation distance. (a) Stage II. (b) Stage III. Plots (a) and (b) share the same color scale.

### Discussion

Very interestingly, MI and frequency comb generation in a cavity which contains purely second-order nonlinear medium have been demonstrated recently [57],[29],[30]. It should be noted, however, that the mechanism for comb initiation in our paper is different from that in Refs. [57],[29],[30]. In particular, in our setup the second-harmonic wave is relatively weak compared to the fundamental wave, such that the comb formation is still dominated by four-wave-mixing due to the third-order Kerr effect. The role of the second-order nonlinearity is to induce phase-shifts around the fundamental frequency (through second-harmonic mode coupling), which enables phase-matching for four-wave mixing. In this sense, the second-order nonlinearity can be regarded as a perturbation to the general single L-L equation. In future, it is also highly interesting to pursue comb generation with much stronger second-order nonlinearity as has been partially investigated in numerical simulations [58].

Currently the fundamental comb bandwidth we demonstrate is far less than one octave; thus external broadening of the comb spectrum will be required for comb self-referencing [59]. Direct octave-spanning comb generation may be achieved by reducing the microresonator dispersion for the fundamental wave. Another highly interesting option is increasing the bandwidths of both the fundamental comb and the second-harmonic comb such that their spectra meet somewhere in the middle. This may allow new freedom in the microresonator dispersion engineering.

### CONCLUSIONS

In conclusion, we have showed that the interaction between the fundamental and the second-harmonic waves can provide a new way of phase-matching for four-wave mixing in the normal-dispersion region, under conditions where comb formation is prohibited in the absence of mode interaction. By employing this new mechanism, we have demonstrated second-harmonic assisted mode-locked comb generation in an on-chip microresonator which shows normal group velocity dispersion for the fundamental wave. A theoretical model incorporating coupled time-domain mean-field equations is developed; and the numerical simulations unequivocally reveal the comb formation process triggered by the fundamental−second-harmonic interactions. Simultaneous comb formation and second-harmonic generation overcoming the dispersion limit would prove especially important in the near-visible to visible range where several atomic transitions commonly used for stabilization of optical clocks are located [60] and where the strong normal material dispersion is likely to dominate. We expect our findings will have a wide impact on the study of nonlinear cavity dynamics with interacting second-order and third-order nonlinearities.

## ACKNOWLEDGEMENTS

This work was supported in part by the National Science Foundation under grant ECCS-1509578, by the Air Force Office of Scientific Research under grant FA9550-15-1-0211, and by the DARPA PULSE program through grant W31P40-13-1-0018 from AMRDEC. Xiaoxiao Xue was supported in part by the National Natural Science Foundation of China under grant 61420106003. Miro Erkintalo acknowledges funding from the Marsden Fund and the Rutherford Discovery Fellowships of the Royal Society of New Zealand.

## AUTHOR CONTRIBUTIONS

X.X. led the experiments, with assistance from J.A.J.V., P.H.W and D.E.L. X.X. and Y.X. designed the SiN microring layout. Y.X. fabricated the microring. X.X., F.L. and M.E. performed the theoretical analysis and did the numerical simulations. X.X., F.L., M.E. and A.M.W. prepared the manuscript. The project was organized and coordinated by A.M.W. and M.Q.

## REFERENCES

[1] Udem Th, Holzwarth R, Hänsch TW. Optical frequency metrology. *Nature* 2002; **416**: 233−237.
[2] Del'Haye P, Schliesser A, Arcizet O, Wilken T, Holzwarth R *et al.* Optical frequency comb generation from a monolithic microresonator. *Nature* 2007; **450:** 1214−1217.
[3] Savchenkov AA, Matsko AB, Ilchenko VS, Solomatine I, Seidel D *et al.* Tunable optical frequency comb with a crystalline whispering gallery mode resonator. *Phys Rev Lett* 2008; **101:** 093902.
[4] Levy JS, Gondarenko A, Foster MA, Turner-Foster AC, Gaeta AL *et al.* CMOS-compatible multiple-wavelength oscillator for on-chip optical interconnects. *Nature Photon* 2010; **4:** 37–40.
[5] Razzari L, Duchesne D, Ferrera M, Morandotti R, Chu S *et al.* CMOS-compatible integrated optical hyper-parametric oscillator. *Nature Photon* 2010; **4:** 41–45.
[6] Grudinin IS, Baumgartel L, Yu N. Frequency comb from a microresonator with engineered spectrum. *Opt Express* 2012; **20:** 6604–6609.
[7] Jung H, Xiong C, Fong KY, Zhang X, Tang HX. Optical frequency comb generation from aluminum nitride microring resonator. *Opt Lett* 2013; **38:** 2810–2813.
[8] Hausmann BJM, Bulu I, Venkataraman V, Deotare P, Lončar M. Diamond nonlinear photonics. *Nature Photon* 2014; **8:** 369–374.
[9] Griffith AG, Lau RKW, Cardenas J, Okawachi Y, Mohanty A *et al.* Silicon-chip mid-infrared frequency comb generation. *Nature Comm* 2015; **6:** 6299.
[10] Kippenberg TJ, Holzwarth R, Diddams SA. Microresonator-based optical frequency combs. *Science* 2011; **332:** 555−559.
[11] Matsko AB, Liang W, Savchenkov AA, Maleki L. Chaotic dynamics of frequency combs generated with continuously pumped nonlinear microresonators. *Opt Lett* 2013; **38:** 525−527.

[12] Ferdous F, Miao H, Leaird DE, Srinivasan K, Wang J *et al.* Spectral line-by-line pulse shaping of on-chip microresonator frequency combs. *Nature Photon* 2011; **5:** 770−776.

[13] Papp SB, Diddams SA. Spectral and temporal characterization of a fused-quartz-microresonator optical frequency comb. *Phys Rev A* 2011; **84:** 053833.

[14] Herr T, Hartinger K, Riemensberger J, Wang CY, Gavartin E *et al.* Universal formation dynamics and noise of Kerr-frequency combs in microresonators. *Nature Photon* 2012; **6:** 480−487.

[15] Saha K, Okawachi Y, Shim B, Levy JS, Salem R *et al.* Modelocking and femtosecond pulse generation in chip-based frequency combs. *Opt Express* 2013; **21:** 1335−1343.

[16] Del'Haye P, Beha K, Papp SB, Diddams SA. Self-injection locking and phase-locked states in microresonator-based optical frequency combs. *Phys Rev Lett* 2014; **112:** 043905.

[17] Huang SW, Zhou H, Yang J, McMillan JF, Matsko A *et al.* Mode-locked ultrashort pulse generation from on-chip normal dispersion microresonators. *Phys Rev Lett* 2015; **114:** 053901.

[18] Herr T, Brasch V, Jost JD, Wang CY, Kondratiev NM *et al.* Temporal solitons in optical microresonators. *Nature Photon* 2014; **8:** 145−152.

[19] Yi X, Yang QF, Yang KY, Suh MG, Vahala K. Soliton frequency comb at microwave rates in a high-Q silica microresonator. *Optica* 2015; **2:** 1078−1085.

[20] Xue X, Xuan Y, Liu Y, Wang PH, Chen S *et al.* Mode-locked dark pulse Kerr combs in normal-dispersion microresonators. *Nature Photon* 2015; **9:** 594−600.

[21] Coillet A, Balakireva I, Henriet R, Saleh K, Larger L *et al.* Azimuthal Turing patterns, bright and dark cavity solitons in Kerr combs generated with whispering-gallery-mode resonators. *IEEE Photon J* 2013; **5:** 6100409.

[22] Liang W, Savchenkov AA, Ilchenko VS, Eliyahu D, Seidel D, et al. Generation of a coherent near-infrared Kerr frequency comb in a monolithic microresonator with normal GVD. *Opt Lett* 2014; **39:** 2920−2923.

[23] Del'Haye P, Coillet A, Loh W, Beha K, Papp SB *et al.* Phase steps and resonator detuning measurements in microresonator frequency combs. *Nature Comm* 2015; **6:** 5668.

[24] Brasch V, Geiselmann M, Herr T, Lihachev G, Pfeiffer MHP *et al.* Photonic chip–based optical frequency comb using soliton Cherenkov radiation. *Science* 2016; **351:** 357−360.

[25] Miller S, Luke K, Okawachi Y, Cardenas J, Gaeta AL *et al.* On-chip frequency comb generation at visible wavelengths via simultaneous second- and third-order optical nonlinearities. *Opt Express* 2014; **22:** 26517−26525.

[26] Jung H, Stoll R, Guo X, Fischer D, Tang HX. Green, red, and IR frequency comb line generation from single IR pump in AlN microring resonator. *Optica* 2014; **1:** 396−399.

[27] Haelterman M, Trillo S, Wabnitz S. Dissipative modulation instability in a nonlinear dispersive ring cavity. *Opt Commun* 1992; **91:** 401–407.

[28] Coen S, Randle HG, Sylvestre T, Erkintalo M. Modeling of octavespanning Kerr frequency combs using a generalized mean-field Lugiato–Lefever model. *Opt Lett* 2013; **38:** 37–39.

[29] Leo F, Hansson T, Ricciardi I, De Rosa M, Coen S *et al.* Walk-off-induced modulation instability, temporal pattern formation, and frequency comb generation in cavity-enhanced second-harmonic generation. *Phys Rev Lett* 2016; **116:** 033901.

[30] Leo F, Hansson T, Ricciardi I, De Rosa M, Coen S *et al.* Frequency comb formation in doubly resonant second-harmonic generation. *Phys Rev A* 2016; **93:** 043831.

[31] Lettieri S, Di Finizio S, Maddalena P, Ballarini V, Giorgis F. Second-harmonic generation in amorphous silicon nitride microcavities. *Appl Phys Lett* 2002; **81:** 4706.

[32] Levy JS, Foster MA, Gaeta AL, Lipson M. Harmonic generation in silicon nitride ring resonators. *Opt Express* 2011; **19:** 11415−11421.

[33] Ning T, Pietarinen H, Hyvärinen O, Simonen J, Genty G *et al.* Strong second-harmonic generation in silicon nitride films. *Appl Phys Lett* 2012; **100:** 161902.

[34] Ning T, Pietarinen H, Hyvärinen O, Kumar R, Kaplas T *et al.* Efficient second-harmonic generation in silicon nitride resonant waveguide gratings. *Opt Lett* 2012; **37:** 4269−4271.

[35] Kitao A, Imakita K, Kawamura I, Fujii M. An investigation into second harmonic generation by Si-rich SiNx thin films deposited by RF sputtering over a wide range of Si concentrations. *J Phys D: Appl Phys* 2014; **47:** 215101.

[36] Carmon T, Yang L, Vahala KJ. Dynamical thermal behavior and thermal self-stability of microcavities. *Opt Express* 2004; **12:** 4742–4750.

[37] Gorodetsky ML, Ilchenko VS. Thermal nonlinear effects in optical whispering-gallery microresonators. *Laser Phys* 1992; **2:** 1004–1009.

[38] Fomin AE, Gorodetsky ML, Grudinin IS, Ilchenko VS. Nonstationary nonlinear effects in optical microspheres. *J Opt Soc Am B* 2005; **22:** 459–465.

[39] Johnson TJ, Borselli M, Painter O. Self-induced optical modulation of the transmission through a high-Q silicon microdisk resonator. *Opt Express* 2006; **14:** 817–831.

[40] He L, Xiao YF, Zhu J, Ozdemir SK, Yang L. Oscillatory thermal dynamics in high-Q PDMS-coated silica toroidal microresonators. *Opt Express* 2009; **17:** 9571–9581.

[41] Baker C, Stapfner S, Parrain D, Ducci D, Leo G *et al.* Optical instability and self-pulsing in silicon nitride whispering gallery resonators. *Opt Express* 2012; **20:** 29076–29089.

[42] Zhang L, Fei Y, Cao Y, Lei X, Chen S. Experimental observations of thermo-optical bistability and self-pulsation in silicon microring resonators. *J Opt Soc Am B* 2014; **31:** 201–206.

[43] Xue X, Xuan Y, Wang C, Wang PH, Liu Y *et al.* Thermal tuning of Kerr frequency combs in silicon nitride microring resonators. *Opt Express* 2016; **24:** 687–698.

[44] Weiner AM. *Ultrafast Optics*. Wiley, 2009.

[45] Savchenkov AA, Matsko AB, Liang W, Ilchenko VS, Seidel D *et al.* Kerr frequency comb generation in overmoded resonators. *Opt Express* 2012; **20:** 27290–27298.

[46] Liu Y, Xuan Y, Xue X, Wang PH, Chen S *et al.* Investigation of mode coupling in normal-dispersion silicon nitride microresonators for Kerr frequency comb generation. *Optica* 2014; **1:** 137–144.

[47] Xue X, Xuan Y, Wang PH, Liu Y, Leaird DE *et al.* Normal-dispersion microcombs enabled by controllable mode interactions. *Laser Photon Rev* 2015; **9:** L23–L28.

[48] Matsko AB, Maleki L. Feshbach resonances in Kerr frequency combs. *Phys Rev A* 2015; **91:** 013831.

[49] Matsko AB, Maleki L. Noise conversion in Kerr comb RF photonic oscillators. *J Opt Soc Am B* 2015; **32:** 232–240.

[50] Coen S, Haelterman M. Modulational instability induced by cavity boundary conditions in a normally dispersive optical fiber. *Phys Rev Lett* 1997; **79:** 4139–4142.

[51] Coen S, Haelterman M, Emplit P, Delage L, Simohamed LM, *et al.* Bistable switching induced by modulational instability in a normally dispersive all-fibre ring cavity. *J Opt B: Quantum Semiclass Opt* 1999; **1:** 36–42.

[52] Savchenkov AA, Rubiola1 E, Matsko AB, Ilchenko VS, Maleki L. Phase noise of whispering gallery photonic hyper-parametric microwave oscillators. *Opt Express* 2008; **16:** 4130–4144.

[53] Hansson T, Modotto D, Wabnitz S. Dynamics of the modulational instability in microresonator frequency combs. *Phys Rev A* 2013; **88:** 023819.

[54] Godey C, Balakireva IV, Coillet A, Chembo YK. Stability analysis of the spatiotemporal Lugiato-Lefever model for Kerr optical frequency combs in the anomalous and normal dispersion regimes. *Phys Rev A* 2014; **89:** 063814.

[55] Sylvestre T, Coen S, Emplit P, Haelterman M. Self-induced modulational instability laser revisited: normal dispersion and dark-pulse train generation. *Opt Lett* 2002; **27:** 482–484.

[56] Del'Haye P, Papp SB, Diddams SA. Hybrid electro-optically modulated microcombs. *Phys Rev Lett* 2012; **109:** 263901.

[57] Ricciardi I, Mosca S, Parisi M, Maddaloni P, Santamaria L *et al.* Frequency comb generation in quadratic nonlinear media. *Phys Rev A* 2015; **91:** 063839.

[58] Hansson T, Leo F, Erkintalo M, Anthony J, Coen S *et al.* Single envelope equation modeling of multi-octave comb arrays in microresonators with quadratic and cubic nonlinearities. *J Opt Soc Am B* 2016; **33:** 1207–1215.

[59] Del'Haye P, Coillet A, Fortier T, Beha K, Cole DC *et al.* Phase-coherent microwave-to-optical link with a self-referenced microcomb. *Nature Photon* 2016; **10:** 516–520.

[60] Ludlow AD, Boyd MM, Ye J. Optical atomic clocks. *Rev Mod Phys* 2015; **87:** 637–701.

# Supplementary Information to
# Second-harmonic assisted four-wave mixing in chip-based microresonator frequency comb generation

Xiaoxiao Xue[1,2*], François Leo[3,4], Yi Xuan[2,5], Jose A. Jaramillo-Villegas[2,6], Pei-Hsun Wang[2], Daniel E. Leaird[2], Miro Erkintalo[4], Minghao Qi[2,5], and Andrew M. Weiner[2,5†]

[1]Department of Electronic Engineering, Tsinghua University, Beijing 100084, China
[2]School of Electrical and Computer Engineering, Purdue University, 465 Northwestern Avenue, West Lafayette, Indiana 47907-2035, USA
[3]The Dodd-Walls Centre for Photonic and Quantum Technologies, Department of Physics, The University of Auckland, Auckland 1142, New Zealand
[4]OPERA-photonics, Université libre de Bruxelles (U.L.B.), 50 Avenue F. D. Roosevelt, CP 194/5, B-1050 Bruxelles, Belgium
[5]Birck Nanotechnology Center, Purdue University, 1205 West State Street, West Lafayette, Indiana 47907, USA
[6]Facultad de Ingenierías, Universidad Tecnológica de Pereira, Pereira, RIS 660003, Colombia
*xuexx@tsinghua.edu.cn; †amw@purdue.edu

## 1. Comb intensity noise

Figure S1 shows the evolution of the comb intensity noise with the pump detuning. The intensity noise is very low in most part of the pump scanning process, except in a small region right before transitioning to detuning stage III where an increased intensity noise is observed (Fig. S1(b) stage II+). The comb spectrum in the noisy region (Fig. S1(c)) is slightly wider than that at stage II (Fig. 2(b) in the main paper). The noisy region is also observed in numerical simulations; see the transmission curves in Fig. 5(a) in the main paper. Figure S2(a) shows the evolution of the intracavity waveform with propagation distance in this region. It can be seen that the optical field becomes unstable. Figure S2(b) shows the averaged comb spectrum (as observed in experiments with a general slow-response optical spectrum analyzer) which is slightly wider than that of the noise-free comb at stage II (Fig. 5(b) in the main paper).

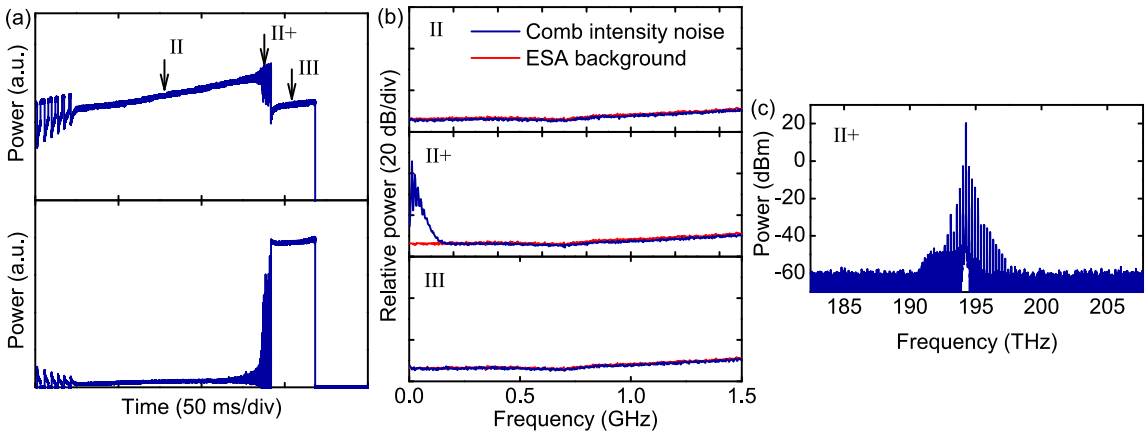

**Figure S1** Evolution of the comb intensity noise. The pump laser scans from shorter wavelength to longer wavelength at a speed of 0.5 nm/s. (a) Drop-port total power (upper) and comb power excluding the pump (lower), measured with the oscilloscope working in sampling mode. An increased intensity noise can be observed at detuning stage II+. (b) Comb intensity noise at different stages marked in (a). ESA: Electrical spectrum analyzer. (c) Fundamental comb spectrum at stage II+. The comb spectra at stages II and III are shown in Fig.2 in the main paper.

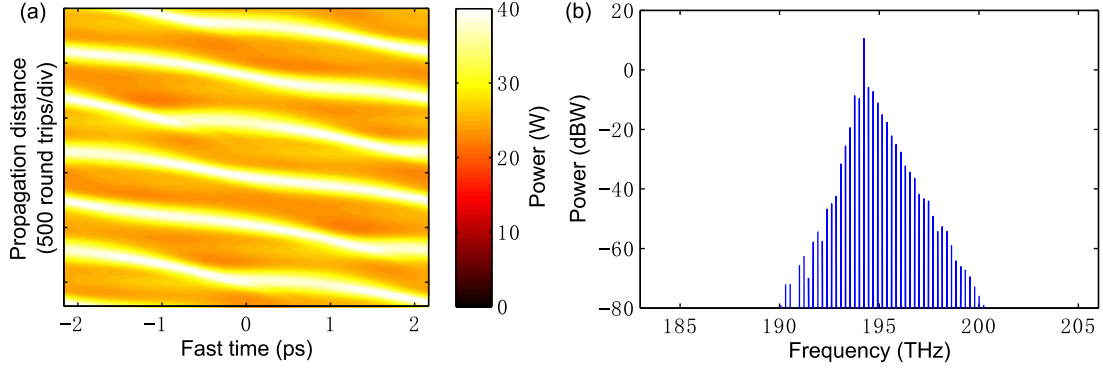

**Figure S2** Simulated comb in the noisy region right before transitioning to stage III. (a) Evolution of the intracavity waveform with propagation distance. (b) Averaged comb spectrum.

## 2. Coupled Lugiato–Lefever (L-L) equations

### 2.1. Equation derivation

The fundamental and second-harmonic waves in the microresonator satisfy the following coupled equations [S1]−[S5]

$$\frac{\partial E_1}{\partial z} = \left[-\alpha_{i1} - i\frac{k_1^{"}}{2}\frac{\partial^2}{\partial \tau^2} + i\gamma_1|E_1|^2 + i2\gamma_{12}|E_2|^2 - \frac{\theta_1}{2}\sum_{n=-\infty}^{+\infty}\delta(z-nL)\right]E_1 + i\kappa E_2 E_1^* e^{-i\Delta k z}$$
$$+ \sum_{n=-\infty}^{+\infty}\delta(z-nL)\sqrt{\theta}E_{in}e^{i\delta_0 z/L} \quad (S1)$$

$$\frac{\partial E_2}{\partial z} = \left[-\alpha_{i2} - \Delta k'\frac{\partial}{\partial \tau} - i\frac{k_2^{"}}{2}\frac{\partial^2}{\partial \tau^2} + i\gamma_2|E_2|^2 + i2\gamma_{21}|E_1|^2 - \frac{\theta_2}{2}\sum_{n=-\infty}^{+\infty}\delta(z-nL)\right]E_2 + i\kappa^* E_1^2 e^{i\Delta k z} \quad (S2)$$

Here, $E_1$ and $E_2$ are amplitude of the fundamental and second-harmonic waves scaled such that $|E_{1,2}|^2$ represents the power flow; $z$ is propagation distance in the cavity; $\tau$ is time; $\alpha_{i1}$ and $\alpha_{i2}$ are amplitude loss per unit length; $k_1^{"} = d^2k/d\omega^2\big|_{\omega=\omega_p}$, $k_2^{"} = d^2k/d\omega^2\big|_{\omega=2\omega_p}$ group velocity dispersion; $\gamma_1$, $\gamma_2$ nonlinear coefficients of self-phase modulation; $\gamma_{12}$, $\gamma_{21}$ nonlinear coefficients of cross-phase modulation; $\theta_1$, $\theta_2$ waveguide-resonator power coupling ratio; $\delta(\cdots)$ delta function; $L$ cavity round-trip length; $\Delta k = 2k(\omega_p) - k(2\omega_p)$ wave vector mismatch; $\Delta k' = dk/d\omega\big|_{\omega=2\omega_p} - dk/d\omega\big|_{\omega=\omega_p}$ group velocity mismatch; $\kappa$ second-order coupling coefficient; $E_{in}$ amplitude of the pump; $\delta_0 = (\omega_0 - \omega_p)t_R$ phase detuning where $\omega_0$ resonance frequency, $\omega_p$ pump frequency, $t_R$ round-trip time for the fundamental wave.

Note that the cavity boundary conditions are here represented by the discrete terms with delta functions. When the change of the intracavity field in one round-trip is very small (meaning that the cavity circumference is much smaller than the nonlinear length and dispersive length, and the cavity loss and pump detuning are very small), the discrete delta functions can be approximated by averaged continuous terms. Thus we get the following mean-field equations

$$\frac{\partial E_1}{\partial z} = \left[-\alpha_1 - i\frac{k_1''}{2}\frac{\partial^2}{\partial \tau^2} + i\gamma_1 |E_1|^2 + i2\gamma_{12}|E_2|^2\right]E_1 + i\kappa E_2 E_1^* e^{-i\Delta k z} + \eta_1 E_{in} e^{i\delta_1 z} \quad (S3)$$

$$\frac{\partial E_2}{\partial z} = \left[-\alpha_2 - \Delta k' \frac{\partial}{\partial \tau} - i\frac{k_2''}{2}\frac{\partial^2}{\partial \tau^2} + i\gamma_2 |E_2|^2 + i2\gamma_{21}|E_1|^2\right]E_2 + i\kappa^* E_1^2 e^{i\Delta k z} \quad (S4)$$

where $\delta_1 = \delta_0/L$, $\alpha_1 = \alpha_{i1} + \theta_1/(2L)$, $\alpha_2 = \alpha_{i2} + \theta_2/(2L)$, $\eta_1 = \sqrt{\theta_1}/L$. Then we do the following variable replacements.

With $E_1' = E_1 e^{-i\delta_1 z}$, $E_2' = E_2 e^{-i2\delta_1 z}$, we get

$$\frac{\partial E_1'}{\partial z} = \left[-\alpha_1 - i\delta_1 - i\frac{k_1''}{2}\frac{\partial^2}{\partial \tau^2} + i\gamma_1 |E_1'|^2 + i2\gamma_{12}|E_2'|^2\right]E_1' + i\kappa E_2' E_1'^* e^{-i\Delta k z} + \eta_1 E_{in} \quad (S5)$$

$$\frac{\partial E_2'}{\partial z} = \left[-\alpha_2 - i2\delta_1 - \Delta k' \frac{\partial}{\partial \tau} - i\frac{k_2''}{2}\frac{\partial^2}{\partial \tau^2} + i\gamma_2 |E_2'|^2 + i2\gamma_{21}|E_1'|^2\right]E_2' + i\kappa^* E_1'^2 e^{i\Delta k z} \quad (S6)$$

With $E_2'' = E_2' e^{-i\Delta k z}$, we get

$$\frac{\partial E_1'}{\partial z} = \left[-\alpha_1 - i\delta_1 - i\frac{k_1''}{2}\frac{\partial^2}{\partial \tau^2} + i\gamma_1 |E_1'|^2 + i2\gamma_{12}|E_2''|^2\right]E_1' + i\kappa E_2'' E_1'^* + \eta_1 E_{in} \quad (S7)$$

$$\frac{\partial E_2''}{\partial z} = \left[-\alpha_2 - i\Delta k - i2\delta_1 - \Delta k' \frac{\partial}{\partial \tau} - i\frac{k_2''}{2}\frac{\partial^2}{\partial \tau^2} + i\gamma_2 |E_2''|^2 + i2\gamma_{21}|E_1'|^2\right]E_2'' + i\kappa^* E_1'^2 \quad (S8)$$

To simply the denotations, we still use $E_1$ and $E_2$ to represent the field amplitude in Eqs. (S7) and (S8). Then we get the mean-field coupled equations in the final form as follows

$$\frac{\partial E_1}{\partial z} = \left[-\alpha_1 - i\delta_1 - i\frac{k_1''}{2}\frac{\partial^2}{\partial \tau^2} + i\gamma_1 |E_1|^2 + i2\gamma_{12}|E_2|^2\right]E_1 + i\kappa E_2 E_1^* + \eta_1 E_{in} \quad (S9)$$

$$\frac{\partial E_2}{\partial z} = \left[-\alpha_2 - i\Delta k - i2\delta_1 - \Delta k' \frac{\partial}{\partial \tau} - i\frac{k_2''}{2}\frac{\partial^2}{\partial \tau^2} + i\gamma_2 |E_2|^2 + i2\gamma_{21}|E_1|^2\right]E_2 + i\kappa^* E_1^2 \quad (S10)$$

## 2.2. Modulational instability due to second-harmonic mode coupling

Modulational instability (MI) analysis can be performed by applying a weak single-frequency perturbation to the continuous-wave (CW) solution of Eqs. (S9) and (S10) and investigating how the perturbation amplitude evolves with the propagation distance. In frequency domain, the perturbation corresponds to two modulation sidebands, i.e. upper and lower sidebands ($\omega_p \pm \omega$ where $\omega$ is the perturbation angular frequency). It is not easy to get analytic solutions of the MI gain for the coupled L-L equations. Here we first qualitatively explain how MI can be enabled by second-harmonic mode coupling in the normal-dispersion region, and then quantitatively calculate the MI gain through numerical simulations. We focus on the situation in which the pump frequency and its second-harmonic line have some phase mismatch; thus the CW second-harmonic power is relatively weak compared to the pump. In this case, the second-harmonic mode coupling can be regarded as a perturbation to the general L-L equation that the fundamental wave follows; and the Kerr nonlinearity induced mixing between the second-harmonic modes can be neglected. The upper (lower) sideband of the fundamental wave is coupled to the upper (lower) sideband of the second-harmonic wave through sum/difference frequency generation (see Fig. S3), i.e.

$$\omega_p + (\omega_p \pm \omega) \to 2\omega_p \pm \omega \text{ and } (2\omega_p \pm \omega) - \omega_p \to \omega_p \pm \omega.$$

When this process is close to phase-matched, an additional phase shift with the propagation distance will be introduced to the upper (lower) sideband mode of the fundamental wave due to the substantial photon exchange in sum/difference frequency generation. This effect is similar to that of linear mode coupling which causes mode splitting, meaning that the sideband mode involved is split to two compound modes with wave vectors different than that of the original mode without second-harmonic coupling. Then it is possible that one of the compound modes, the pump and the other sideband mode satisfy an equivalent anomalous dispersion, i.e. $k'_{\omega_p+\omega} - k_{\omega_p} - (k_{\omega_p} - k_{\omega_p-\omega}) < 0$ (here we suppose the upper sidebands are close to phase-matched; $k'_{\omega_p+\omega}$ represents the wave vector of the compound mode involved). MI will thus be enabled.

For the simulation results shown in Figs. 5 and 6 in the main paper, the parameters are as follows: $\alpha_1 = 4.94 \text{ m}^{-1}$, $k_1'' = 186.9 \text{ ps}^2\text{km}^{-1}$, $\gamma_1 = 0.9 \text{ m}^{-1}\text{W}^{-1}$, $\eta_1 = 3.07 \text{ m}^{-1}$, $\alpha_2 = 9.87 \text{ m}^{-1}$, $k_2'' = -151.6 \text{ ps}^2\text{km}^{-1}$, $\gamma_2 = 2.6 \text{ m}^{-1}\text{W}^{-1}$, $\Delta k = -854.66 \text{ m}^{-1}$, $\Delta k' = 6.17 \times 10^{-10} \text{ s} \cdot \text{m}^{-1}$, $\gamma_{12} = 0.72 \text{ m}^{-1}\text{W}^{-1}$, $\gamma_{21} = 1.44 \text{ m}^{-1}\text{W}^{-1}$, $\kappa = 1.3 \text{ m}^{-1}\text{W}^{-1/2}$, $FSR = 231.3 \text{ GHz}$, $E_{\text{in}} = 0.447 \text{ W}^{1/2}$. The propagation loss for the fundamental wave is extracted from the measured cavity quality factor in 1550-nm range. Due to the lack of a proper tunable laser source in 750-nm range, we cannot measure the quality factor for the second-harmonic wave. Thus we assume the loaded quality factor for the second-harmonic wave is half of that for the fundamental wave. The spatial mode for the fundamental wave is the fundamental transverse electric (TE) mode which is verified experimentally by measuring the polarization and cavity FSR. The spatial mode for the second-harmonic wave is investigated through numerical simulations using a commercial software *COMSOL*, and is found to be the 4[th] order TE mode which has the minimum phase mismatch with the fundamental wave (the second-harmonic phase matching wavelength is ~1620 nm). The group velocity dispersion for the fundamental wave is measured with frequency comb assisted spectroscopy [S6], while for the second-harmonic wave is obtained through numerical simulations. The self- and cross-phase modulation coefficients are calculated based on the simulated mode distributions. The nonlinear refractive index is $n_2 = 2.4 \times 10^{-19}$ m$^2$W$^{-1}$ [S7]. Since the wave vector mismatch is very sensitive to the actual material dispersion, this parameter is freely tuned in our simulations. Another freely tuned parameter is the second-order coupling coefficient which is given by [S8]

$$\kappa = \frac{2\varepsilon_0 \omega_p d_{\text{eff}}}{4} \iint \left[E^{2\omega_p}(x,y)\right]^* \left[E^{\omega_p}(x,y)\right]^2 dxdy \qquad (S11)$$

where $\varepsilon_0$ is the vacuum permittivity; $\omega_p$ is angular frequency of the pump; $E^{2\omega_p}$ and $E^{\omega_p}$ are normalized field distribution such that $\frac{1}{2}\iint E \times H^* dxdy = 1$; $d_{\text{eff}} = \chi^{(2)}_{\text{eff}}/2$ and $\chi^{(2)}_{\text{eff}}$ is in the order of $10^{-12} \sim 10^{-11}$ which is in the range reported in literature [S9]−[S11].

Figure S4 shows the simulated MI gain for detuning stage II in Fig. 5 in the main paper ($\delta_0 = 1.67 \times 10^{-2}$). To run the simulation, a series of low-level probe sidebands is added to the CW fields and the coupled L-L equations are integrated with the split-step Fourier method. The evolution of the sideband amplitude is given by $A = A_0 e^{gL}$ where $g$ is the net gain and $L$ is the propagation distance. A narrow-band region with positive net gain is observed 1 FSR away from the pump, which is consistent with the 1-FSR comb generation observed in both experiments and simulations. The wave vector mismatch for sum/difference frequency generation involving the pump is also plotted. It can be found that the lower-sideband gain area is very close to phase-matched condition. This agrees with the second-harmonic mode coupling mechanism described above. The close-to-phase-match coupling for the lower sideband gives rise to a strong lower-sideband second-harmonic line which can be observed in both experiments and simulations (see Figs. 2(c) and 5(c) in the main paper).

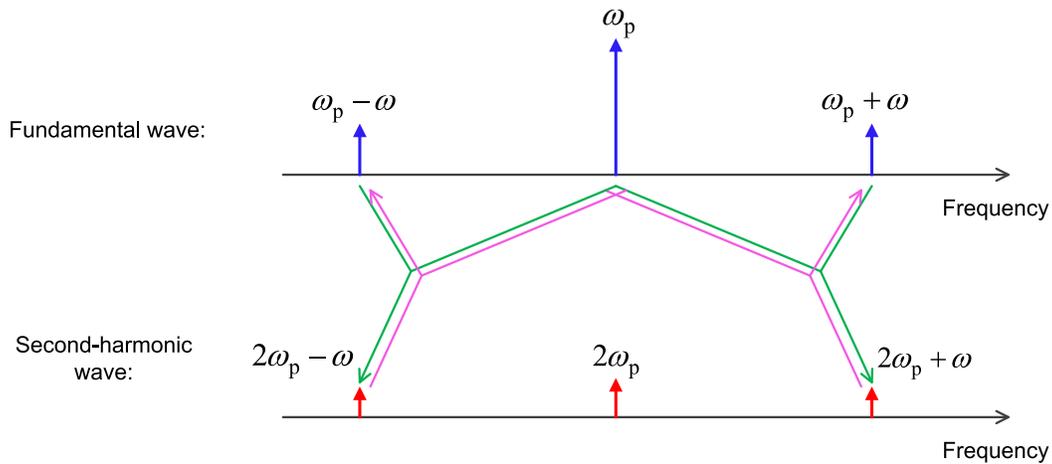

**Figure S3** Illustration of mode coupling through sum/difference frequency generation.

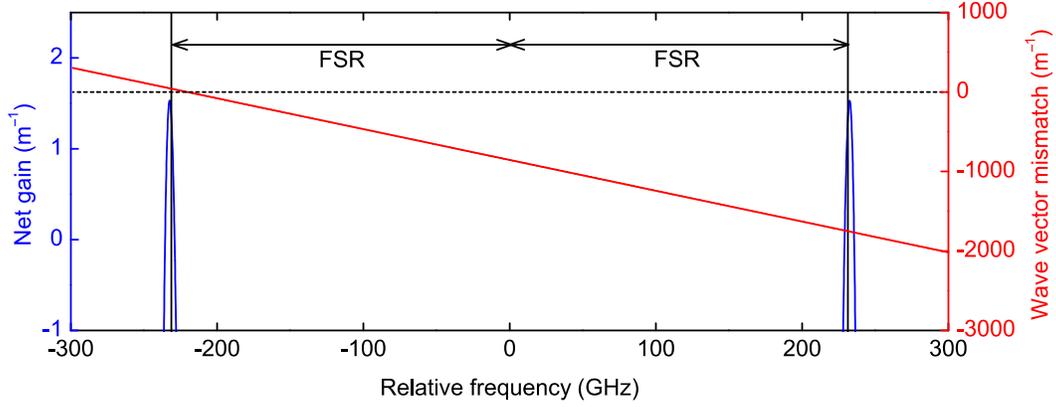

**Figure S4** Numerically simulated MI gain for detuning stage II in Fig. 5 in the main paper. A narrow-band region with positive net gain can be observed 1 FSR away from the pump. The wave vector mismatch for sum/difference frequency generation, i.e. $k(\omega_p) + k(\omega_p + \omega) - k(2\omega_p + \omega) = \Delta k - \Delta k' \omega$, is also plotted. The lower-sideband gain area is very close to phase-matched condition.

### 2.3. Comparison between experimental and simulation results

Figure S5 shows a side-by-side plot of the experimental and simulation results. The comb generation behavior revealed by the simulation agrees well with our experimental observation. With the pump laser frequency tuned from blue to red, a narrow-band fundamental comb is first generated, and then transitions to a state with a much broader spectrum. Evolution of the second-harmonic spectrum can also be observed. Moreover, the change of the comb line spacing accompanied with the comb state switching is also captured by the simulation (see Fig. 6 of the main text). Note that the thermal effect is not considered in the simulation, thus no oscillation is found in the simulated curves. The bandwidth of the simulated broadband comb at stage III is close to that of the experimental measurement. The different fine spectral features are related to different temporal features of the intracavity dark pulse, which are very sensitive to the comb operating parameters [S12]−[S15].

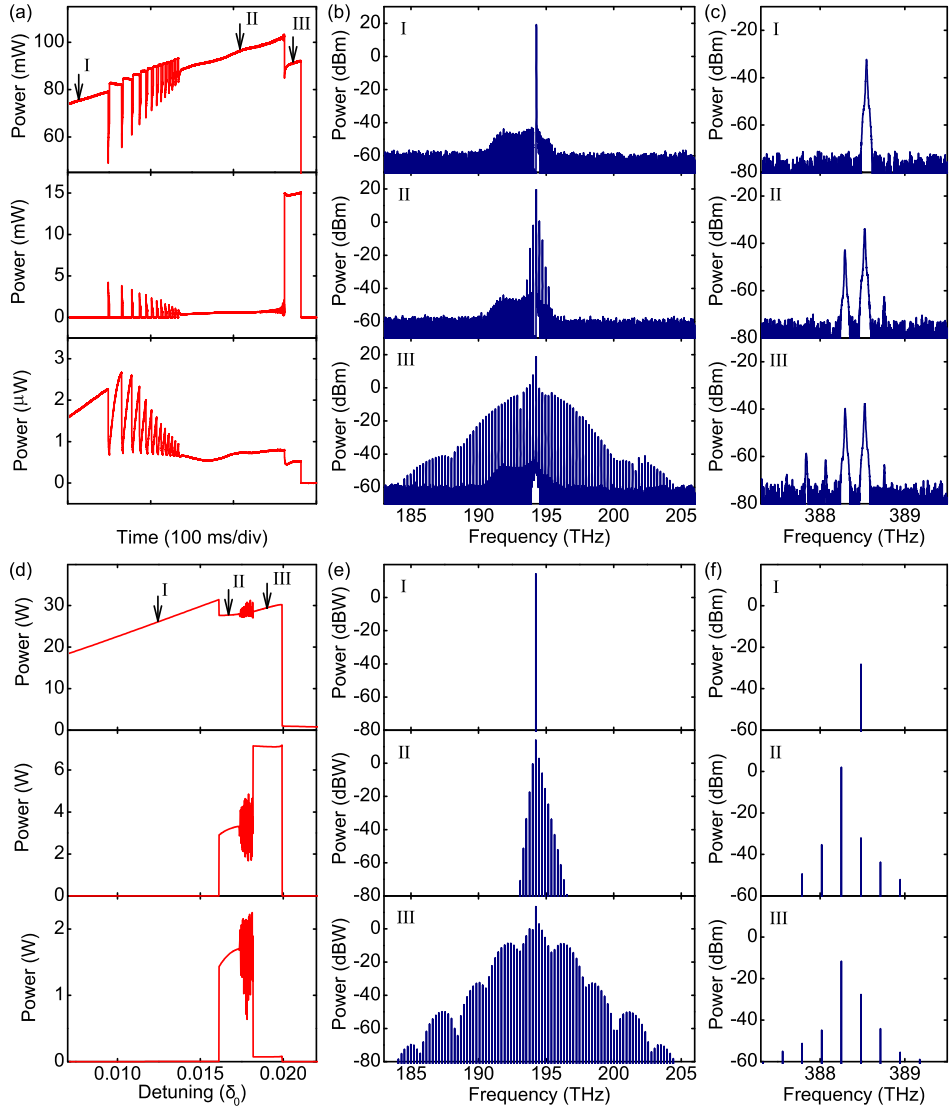

**Figure S5** Side-by-side plot of the experimental (a-c) and simulation (d-f) results. (a, d) Transition curves when the pump laser scans across the resonance from shorter to longer wavelength. From top to bottom: total infrared (IR) power, IR comb power excluding the pump line, second-harmonic power. (b, c) Measured IR and second-harmonic comb spectra at different detuning stages marked in (a). (e, f) Simulated IR and second-harmonic comb spectra at different detuning stages marked in (d).

## References for Supplementary Information

[S1] Agrawal GP. *Nonlinear Fiber Optics*. Academic Press, 2001.

[S2] Menyuk CR, Schiek R, Tornert L. Solitary waves due to χ(2):χ(2) cascading. *J Opt Soc Am B* 1994; **11:** 2434−2443.

[S3] Buryak AV, Trapani PD, Skryabin DV, Trillo S. *Phys Rep* 2012; **370:** 63–235.

[S4] Leo F, Hansson T, Ricciardi I, De Rosa M, Coen S *et al.* Walk-off-induced modulation instability, temporal pattern formation, and frequency comb generation in cavity-enhanced second-harmonic generation. *Phys Rev Lett* 2016; **116:** 033901.

[S5] Leo F, Hansson T, Ricciardi I, De Rosa M, Coen S *et al.* Frequency comb formation in doubly resonant second-harmonic generation. *Phys Rev A* 2016; **93:** 043831.

[S6] Del'Haye P, Arcizet O, Gorodetsky ML, Holzwarth R, Kippenberg TJ. Frequency comb assisted diode laser spectroscopy for measurement of microcavity dispersion. *Nature Photon* 2009; **3:** 529–533.

[S7] Ikeda K, Saperstein RE, Alic N, Fainman Y. Thermal and Kerr nonlinear properties of plasma-deposited silicon nitride/silicon dioxide waveguides. *Opt Express* 2008; **16:** 12987–12994.

[S8] Suhara T, Fujimura M. *Waveguide Nonlinear Optic Devices*. Springer, 2003.

[S9] Ning T, Pietarinen H, Hyvärinen O, Simonen J, Genty G *et al.* Strong second-harmonic generation in silicon nitride films. *Appl Phys Lett* 2012; **100:** 161902.

[S10] Ning T, Pietarinen H, Hyvärinen O, Kumar R, Kaplas T *et al.* Efficient second-harmonic generation in silicon nitride resonant waveguide gratings. *Opt Lett* 2012; **37:** 4269−4271.

[S11] Kitao A, Imakita K, Kawamura I, Fujii M. An investigation into second harmonic generation by Si-rich SiNx thin films deposited by RF sputtering over a wide range of Si concentrations. *J Phys D: Appl Phys* 2014; **47:** 215101.

[S12] Xue X, Xuan Y, Liu Y, Wang PH, Chen S et al. Mode-locked dark pulse Kerr combs in normal-dispersion microresonators. *Nature Photon* 2015; **9:** 594−600.

[S13] Godey C, Balakireva IV, Coillet A, Chembo YK. Stability analysis of the spatiotemporal Lugiato-Lefever model for Kerr optical frequency combs in the anomalous and normal dispersion regimes. *Phys Rev A* 2014; **89:** 063814.

[S14] Parra-Rivas P, Gomila D, Knobloch E, Coen S, Gelens L, Origin and stability of dark pulse Kerr combs in normal dispersion resonators. *Opt Lett* 2016; **41:** 2402−2405.

[S15] Parra-Rivas P, Knobloch E, Gomila D, Gelens L. Dark solitons in the Lugiato-Lefever equation with normal dispersion. *Phys Rev A* 2016; **93:** 063839.